\documentclass[11pt]{article}
\usepackage{fullpage}
\usepackage{graphicx}
\usepackage{latexsym,amsmath,amssymb,amsthm,epic,eepic,multirow}
\usepackage{natbib}

\usepackage{multirow}
\usepackage{multicol}
\usepackage{subcaption}
\usepackage{natbib}
\usepackage{algorithm}
\usepackage{algorithmic}
\usepackage{mdframed}
\usepackage{tikz}
\usetikzlibrary{positioning}
\usepackage{pgf,tikz}
\usepackage{hyperref}
\usepackage{listings} 
\usepackage{color} 
\usepackage{bbm}
\usepackage{relsize}
\usepackage{array}
\usepackage{mathtools}
\newcolumntype{P}[1]{>{\centering\arraybackslash}p{#1}}

\definecolor{upmaroon}{rgb}{0.48, 0.07, 0.07}
\definecolor{royalazure}{rgb}{0.0, 0.22, 0.66}
\definecolor{pakistangreen}{rgb}{0.0, 0.4, 0.0}

\newcommand{\EE}{\mathbb{E}}

\theoremstyle{definition}
\newtheorem{theo}{Theorem}
\newtheorem{prop}{Proposition}

\newtheoremstyle{dotless}{}{}{}{}{\bfseries}{}{ }{}
\theoremstyle{dotless}
\newtheorem{assa}{}

\usepackage{color}

\let\originalleft\left
\let\originalright\right
\renewcommand{\left}{\mathopen{}\mathclose\bgroup\originalleft}
\renewcommand{\right}{\aftergroup\egroup\originalright}
\makeatletter
\newcommand{\leqnomode}{\tagsleft@true}
\newcommand{\reqnomode}{\tagsleft@false}
\makeatother

\begin{document}

\title{On the pitfalls of Gaussian likelihood scoring for causal discovery}
  
\author{Christoph Schultheiss and Peter B\"uhlmann\\
Seminar for Statistics, ETH Z\"urich}

\maketitle

\begin{abstract}
We consider likelihood score-based methods for causal discovery in structural causal models. In particular, we focus on Gaussian scoring and analyze the effect of model misspecification in terms of non-Gaussian error distribution. We present a surprising negative result for Gaussian likelihood scoring in combination with nonparametric regression methods.
\end{abstract}


\section{Introduction}
We consider the problem of finding the causal structure of a set of random variables $X_1, \ldots X_p$. We assume that the data can be represented by a structural equation model whose structure is given by a directed acyclic graph (DAG), say, $G^0$, with nodes $1, \ldots, p$ and denote by $\text{PA}\left(j\right)$ the parents of $j$. 
Without further assumptions on the structural causal model, one can only find $G^0$ up to its Markov equivalence class. This can, e.g., be achieved with the PC algorithm \citep{spirtes2000causation}. Its generality comes at the price of requiring conditional independence tests, which are a statistically hard problem.

Here, we focus on the so-called additive noise model (ANM)
\begin{equation}\label{eq:ANM}
X_j \leftarrow f_j \left(X_{\text{PA}\left(j\right)}\right) + \mathcal{E}_j \quad \forall j \in 1, \ldots, p,
\end{equation}
where the $\mathcal{E}_j$ are mutually independent centered random variables. This is a popular modelling assumption that allows for better identifiability guarantees; see, e.g., \citep{hoyer2008nonlinear, peters2014causal}.

For an arbitrary DAG $G$, let $\text{PA}^G\left(j\right)$ be the nodes from which a directed edge towards $j$ starts. Define
\begin{equation*}
\mathcal{E}^G_j = X_j - \EE\left[X_j \vert X_{\text{PA}^G\left(j\right)} \right] \quad \forall j \in 1, \ldots, p.
\end{equation*}
Obviously, $\mathcal{E}^{G^0}_j=\mathcal{E}_j$. Under not overly restrictive assumptions, it holds that $\mathcal{E}^G_1, \ldots \mathcal{E}^G_p$ are mutually independent only if $G \supseteq G^0$, see \cite{peters2014causal}. Thus, an obvious approach to find $G^0$ is to loop over all possible graphs and test for independence of the residuals. Of course, this becomes infeasible when the dimensionality $p$ grows.

A more reasonable algorithm based on greedy search is presented in \cite{peters2014causal}. They also introduce RESIT (regression with subsequent independence test) that iteratively detects sink nodes. Finding the true DAG is guaranteed assuming perfect regressors and independence tests. It involves $O\left(p^2\right)$ nonparametric independence tests which might be computationally involved or lacking power when the sample size is small.

Instead of performing independence tests, one can compare the likelihood score of different graphs
\begin{equation*}
\mathcal{L}\left(G\right)=\EE\left[\log\left(\prod_{j=1}^pp_j^G\left(\mathcal{E}_j^G\right)\right)\right]=\sum_{j=1}^p \EE\left[\log\left(p_j^G\left(\mathcal{E}_j^G\right)\right)\right],
\end{equation*}
where $p_j^G$ denotes the density of $\mathcal{E}_j^G$. Only for independent $\mathcal{E}_j^G$, their multivariate density factorizes as suggested. Therefore, the true DAG maximizes this quantity due to the properties of the Kullback-Leibler divergence. In practice, this comes with the additional difficulty of estimating the densities $p_j^G\left(\cdot\right)$ and one needs to add some penalization to prefer simpler graphs and avoid selecting complete graphs; see, e.g., \cite{nowzohour2016score}.

If one additionally assumes that the $\mathcal{E}_j$ are marginally normally distributed
$\mathcal{N}\left(0,\sigma_j^2\right)$ one can instead consider the Gaussian likelihood
\begin{equation*}
\mathcal{L}^\mathcal{N}\left(G\right) \coloneqq \sum_{j=1}^p\left(-\log\left(\sigma^G_j\right) - \dfrac{1}{2} - \dfrac{1}{2}\log\left(2 \pi\right)\right)=-\sum_{j=1}^p\log\left(\sigma^G_j\right) + C,
\end{equation*}
where $\left(\sigma_j^G\right)^2 =\EE\left[\left(\mathcal{E}_j^G\right)^2\right]$ \citep{buhlmann2014cam}. It holds $\mathcal{L}\left(G\right) \geq \mathcal{L}^\mathcal{N}\left(G\right)$ and $\mathcal{L}^\mathcal{N}\left(G^0\right) = \mathcal{L}\left(G^0\right)$.
Thus, the problem simplifies to finding the graph that leads to the lowest sum of log-variances. 

Under such a normality assumption for $\mathcal{E}_j$ and the $f_j\left(\cdot\right)$ in \eqref{eq:ANM} being additive in their arguments, \cite{buhlmann2014cam} present a causal discovery method that is consistent for high-dimensional data. 
To justify this approach for a broader class of error distributions, define
\begin{equation}\label{eq:Delta}
\Delta \coloneqq \underset{G \not \supseteq G^0}{\text{min}} \sum_{j=1}^p \log\left(\sigma^G_j\right) - \log\left(\sigma^{G^0}_j\right)
\end{equation}
Then, one has to assume that
\begin{assa}\label{ass:gap}
$\Delta  > 0$.
\end{assa}
That is, the lowest possible expected negative Gaussian log-likelihood with any graph $G$ that is not a superset of the true $G^0$ must be strictly larger than the expected negative Gaussian log-likelihood with the true graph $G^0$.
An argument for that assumption is that a true causal model should be easier to fit in some sense and thus also obtain lower error variance. Using simple ``non-pathological'' examples, we demonstrate that this can easily be a fallacy when the true error distribution is misspecified. Thus, we advocate the need to be very careful when using Gaussian scoring with flexible nonparametric regression functions in causal discovery. The main part of this paper considers theoretical population properties. We discuss some data applications in Section \ref{dat}.

\section{Data-generating linear model}
We consider first data-generating linear models where
\begin{equation*}
f_j\left(X_{PA\left(j\right)}\right) = \sum_{k \in PA\left(j\right)} \beta_{jk} X_k.
\end{equation*}
For these, we find the explicit Theorem \ref{theo:lingam}. The intuition for this result carries over to a range of nonlinear ANM \eqref{eq:ANM}, especially if the causal effects are close to linear. We present according examples in Section \ref{beyond}.

If all the $\mathcal{E}_j$ in a data-generating linear model are Gaussian, i.e., $X_1, \ldots, X_p$ are jointly Gaussian, it is known that any causal order could induce the given multivariate normal distribution and obtain an optimal Gaussian score. Assuming that the distribution is faithful with respect to the true DAG $G^0$, the set of the most sparse DAGs obtaining the optimal Gaussian score corresponds to the Markov equivalence class of $G^0$; see, e.g., \cite{zhang2008detection}. Thus, one can obtain this Markov equivalence class by preferring more sparse DAGs in case of equal scores. In general, the single true DAG cannot be determined even if the full multivariate distribution is known. On the contrary, the Linear Non-Gaussian Acyclic Model (LiNGAM) introduced in \cite{shimizu2006linear} is known to be identifiable. In such linear non-Gaussian models, algorithms designed for linear Gaussian models, e.g., the PC algorithm using partial correlation to assess conditional independence, do not use all the available information, but typically still provide the same guarantees since they depend on the covariance structure only. The covariance matrix of data generated by a linear causal model does not change when replacing a Gaussian $\mathcal{E}_j$ by an additive error of the same variance but otherwise arbitrary distribution. Thus, assuming the faithfulness condition for the data-generating distribution, Gaussian scoring for linear causal models in the infinite sample limit leads to the true underlying Markov equivalence class even under misspecification of the error distribution.

If the data-generating model is not known to be linear such that nonparametric regression methods, or the conditional mean as their population version, are applied, this generalization does not hold true anymore, as laid out in the following theorem.  Let $\pi$ be a permutation on $\left\{1,\ldots,p\right\}$ and $G^\pi$ the full DAG according to $\pi$, i.e., $\pi\left(k\right) \in \text{PA}^{G^\pi}\left(\pi\left(j\right)\right)$ if and only if $k < j$. 

\begin{theo}\label{theo:lingam}
Let $X_1, \ldots, X_p$ come from a linear model:
\begin{equation}\label{eq:linmod}
X_j \leftarrow \sum_{k \in PA\left(j\right)} \beta_{jk} X_k + \mathcal{E}_j, \quad \text{with} \quad \EE\left[\mathcal{E}_j\right]=0, \quad \EE\left[\mathcal{E}_j^2\right]< \infty \quad \forall j \in 1, \ldots, p,
\end{equation}
with mutually independent $\mathcal{E}_j$.
Then, for every possible permutation $\pi$,
\begin{equation*}
\quad \sum_{j=1}^p \log\left(\sigma^{G^\pi}_j\right) - \log\left(\sigma^{G^0}_j\right) \leq 0.
\end{equation*}
That is, for every causal order, the corresponding full graph scores at least as well as the true causal graph.\\
Furthermore, 
\begin{align*}
&\sum_{j=1}^p \log\left(\sigma^{G^\pi}_j\right) - \log\left(\sigma^{G^0}_j\right) < 0 \quad \text{if}\\
&\exists j \in \left\{1,\ldots,p\right\}: \quad \EE\left[X_j \vert X_{PA^{G^\pi}\left(j\right)}\right] \neq  X_{PA^{G^\pi}\left(j\right)}^\top \beta^{\pi,j}\quad \text{where}\\
&\beta^{\pi,j}  =  \EE\left[X_{PA^{G^\pi}\left(j\right)} X_{PA^{G^\pi}\left(j\right)}^\top\right]^{-1}\EE\left[X_j X_{PA^{G^\pi}\left(j\right)}\right] \quad \text{is the least squares parameter}.
\end{align*}
That is, for every causal order, the corresponding full graph scores strictly better than the true causal graph if at least one conditional expectation is nonlinear in the parental variables.
\end{theo}
Apart from some pathological cases, the last condition holds for permutations that are not conformable with the true DAG unless all the $\mathcal{E}_j$ are Gaussian. Thus, the gap condition \ref{ass:gap} does not hold true for this whole set of distributions, and, without Gaussianity, a wrong model would not only score equivalently but even be preferred. This is in stark contrast to results around Gaussian scoring when fitting linear models.

\subsection{Illustrative examples}
For illustrative purposes, we restrict ourselves to the two variable case with $X_2 \leftarrow \beta X_1 + \mathcal{E}_2$. We have the true DAG $G^0 = \left\{X_1 \rightarrow X_2\right\}$ and define $G^\perp = \left\{X_1 \leftarrow X_2\right\}$. If $ \beta X_1 \overset{\mathcal{D}}{=}\mathcal{E}_2$, it holds
\begin{equation*}
X_2 = \EE\left[X_2 \vert X_2\right] = \EE\left[\beta X_1 + \mathcal{E}_2 \vert X_2\right] = 2 * \beta \EE\left[X_1 \vert X_2\right] \quad \text{implying that} \quad \EE\left[X_1 \vert X_2\right]= \dfrac{1}{2 \beta}X_2
\end{equation*}
such that $\Delta = 0$; see the definition in \eqref{eq:Delta}. In the two variable linear model, $\exp\left(\Delta\right)^2$ equals the ratio of the attainable mean squared error for the backward direction between the best-fitting unrestricted model and the best-fitting linear model.

\begin{figure}[b!]
\begin{subfigure}{.33\textwidth}
  \centering
  \includegraphics[width=.8\linewidth]{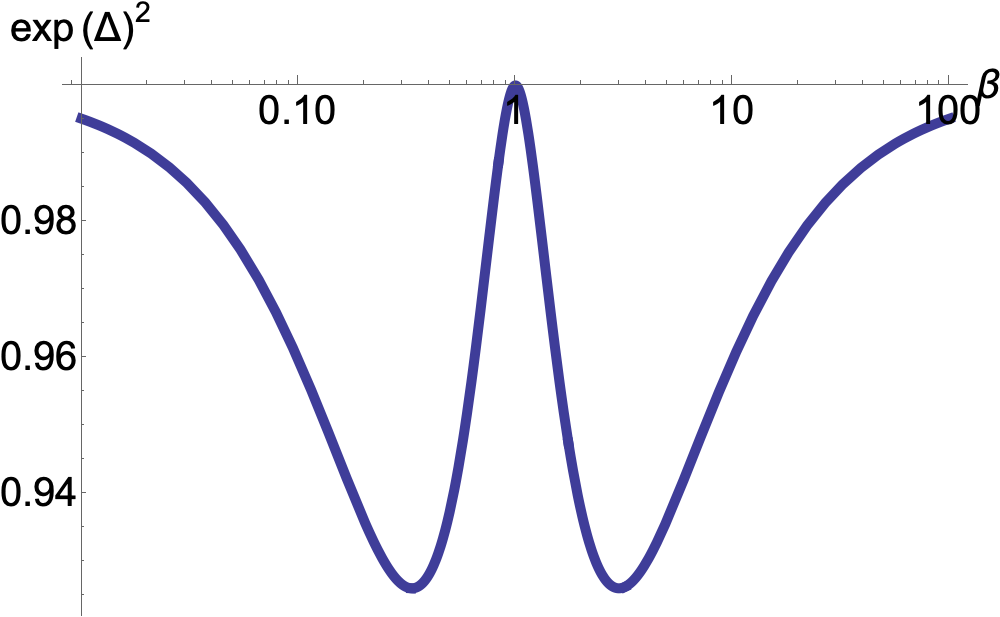}
  \caption{{$X_1 \overset{\mathcal{D}}{=}\mathcal{E}_2 \sim \text{Unif}\left[-1,1\right]$}}
  \label{fig:uni-uni}
\end{subfigure}
\begin{subfigure}{.33\textwidth}
  \centering
  \includegraphics[width=.8\linewidth]{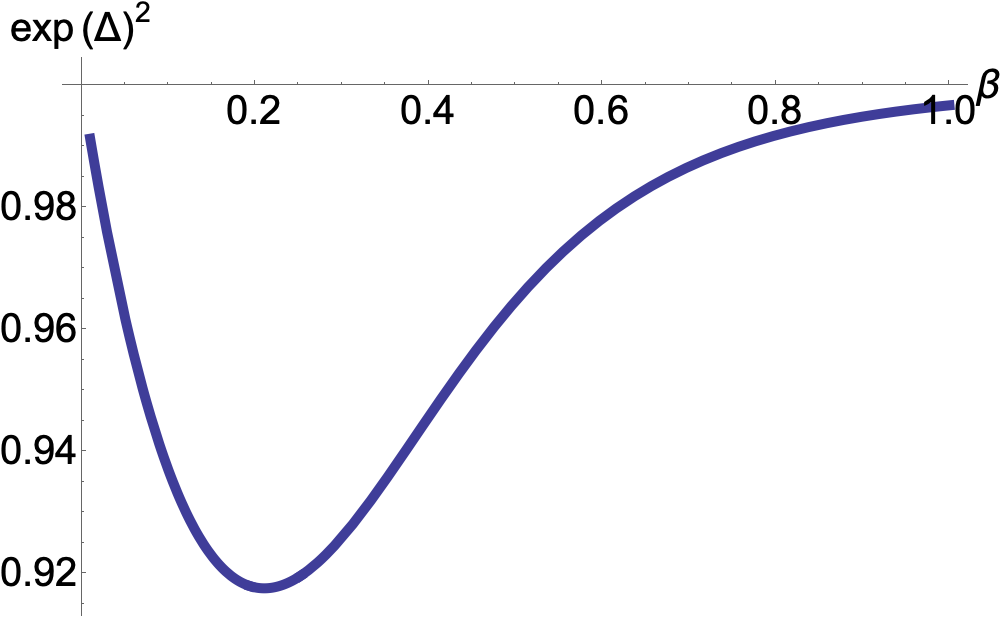}
  \caption{{$X_1 \sim \mathcal{N}\left(0,1\right)$, $\mathcal{E}_2 \sim \text{Unif}\left[-1,1\right]$}}
  \label{fig:ga-uni}
\end{subfigure}
\begin{subfigure}{.33\textwidth}
  \centering
  \includegraphics[width=.8\linewidth]{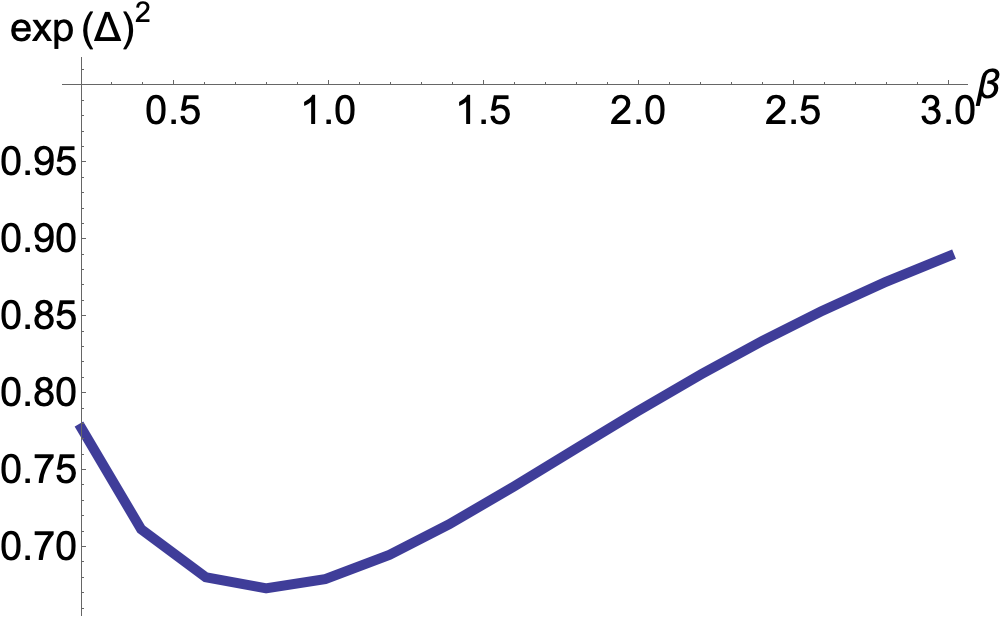}
  \caption{{$X_1 \sim \mathcal{N}\left(0,1\right)$, $\mathcal{E}_2 + 1 \sim \chi_1^2$}}
  \label{fig:ga-xi}
\end{subfigure}
\caption{Two variable linear model: effect of changing $\beta$.}
\label{fig:lin-add}
\end{figure}

Consider first the analytically tractable case where $X_1 \overset{\mathcal{D}}{=}\mathcal{E}_2 \sim \text{Unif}\left[-1,1\right]$. Then, for $\beta= \pm 1$, $\EE\left[X_1 \vert X_2\right]= \pm X_2 / 2$ and $\Delta = 0$. For every other nonzero and bounded $\beta$, $\Delta <0$ so a causal discovery method based on Gaussian scoring would - assuming a correct regression function estimator - wrongly claim $X_2 \rightarrow X_1$ in the large sample limit. We make things more explicit.
\begin{prop}\label{prop:lin-unif}
Let $X_2 = \beta X_1 + \mathcal{E}_2$ with $X_1 \overset{\mathcal{D}}{=}\mathcal{E}_2 \sim \text{Unif}\left[-1,1\right]$. Define $\gamma = \text{max}\left\{\left\vert\beta\right\vert,\left\vert 1 / \beta\right \vert \right\}$. Then, the ratio of the variances is given by
\begin{equation*}
\exp\left(\Delta\right)^2=\left(\prod_{j=1}^2 \left(\sigma_j^{G^\perp}\right)^2\right)/\left(\prod_{j=1}^2 \left(\sigma_j^{G^0}\right)^2\right) = \dfrac{\left(\gamma^2 + 1\right)\left(2 \gamma - 1\right)}{2 \gamma^3} \coloneqq r\left(\gamma\right).
\end{equation*}
\end{prop}
As argued above, $r \left(1\right) = 1$. For $\left\vert\beta\right\vert \rightarrow 0$, the ratio between the variance products approaches $1$ as $X_1$ and $X_2$ become independent, and, hence, both models perform equally well. For $\left\vert\beta\right\vert \rightarrow \infty$, the ratio approaches $1$ as $X_2$ becomes a deterministic linear map of $X_1$, and, hence, the linear model is invertible. The ratio $r\left(\gamma\right)$ is minimized for $\gamma = 3$, with $r\left(3\right) \approx 0.93$, strictly decreasing for $\gamma \in \left[1,3\right)$, and strictly increasing for $\gamma \in \left(3,\infty\right)$. This is visualized in Figure \ref{fig:uni-uni}.

Consider next a similar example but with $X_1 \sim \mathcal{N}\left(0,1\right)$. Analytic expressions for $\EE\left[X_1 \vert X_2 \right]$ and $\text{Var}\left(X_1 \vert X_2\right)$ can be found in terms of the Gaussian cumulative distribution function. For $\left(\sigma_1^{G^\perp}\right)^2 = \EE\left[\text{Var}\left(X_1 \vert X_2\right)\right]$ we invoke numerical integration. The obtained $\exp\left(\Delta\right)^2$ is shown in Figure \ref{fig:ga-uni}. As elaborated before, it approaches $1$ for $\beta \rightarrow 0$ and $\beta \rightarrow \infty$. In between, it is strictly less than $1$ since $\EE\left[X_1 \vert X_2\right]$ is not linear in $X_2$. The minimum of $\exp\left(\Delta\right)^2 \approx 0.92$ is obtained for $\beta \approx 0.21$.

Finally, we use an asymmetric error distribution instead, namely, $\mathcal{E}_2 \sim \chi_1^2 -1$. Figure \ref{fig:ga-xi} shows that this leads to more extreme values of $\Delta < 0$. Notably, $\EE\left[X_1 \vert X_2\right]$ is not monotone in this case so the best linear fit is not a good approximation. 

\section{Beyond a data-generating linear model}\label{beyond}
If all $\mathcal{E}_j$ are Gaussian, we know that $\Delta = 0$ for linear conditional expectations in \eqref{eq:ANM}, but $\Delta > 0$ otherwise. Thus, involving nonlinearities enables the identifiability of the model. 

For non-Gaussian $\mathcal{E}_j$ in a linear model, $\Delta < 0$ holds true apart from some special cases; see, Theorem \ref{theo:lingam}. The intuition is that nonlinearities could be beneficial for the identifiability nevertheless. As the lower bound for $\Delta$ is negative and not $0$, presumably a higher degree of nonlinearity might be necessary to achieve $\Delta > 0$. We analyze this with the following simple model 
\begin{equation}\label{eq:potmod}
X_2 \leftarrow \beta \text{Sign}\left(X_1\right) \dfrac{\left\vert X_1 \right\vert ^\nu}{\sqrt{\EE\left[\left\vert X_1 \right\vert ^{2\nu}\right]}} + \mathcal{E}_2, \quad \nu > 0.
\end{equation}
The normalization ensures that the variance of $X_2$ depends only on $\beta$. For $\nu = 1$, we obtain a linear model.

\begin{figure}[b!]
\begin{subfigure}{.33\textwidth}
  \centering
  \includegraphics[width=.8\linewidth]{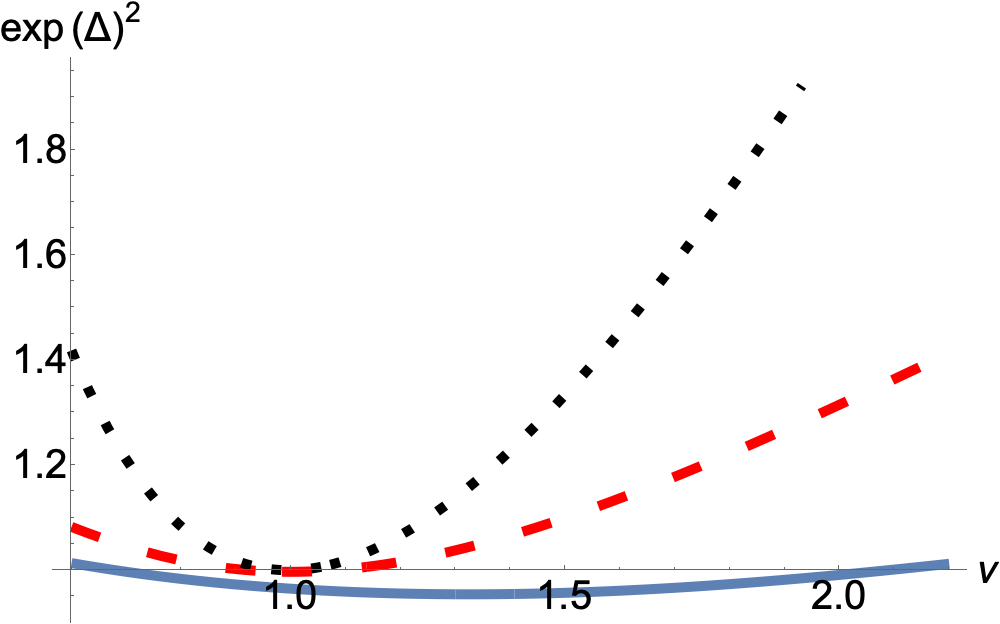}
  \caption{{$X_1 \sim \mathcal{N}\left(0,1\right)$, $\mathcal{E}_2 \sim \text{Unif}\left[-1,1\right]$}}
  \label{fig:ga-uni-pot}
\end{subfigure}
\begin{subfigure}{.33\textwidth}
  \centering
  \includegraphics[width=.8\linewidth]{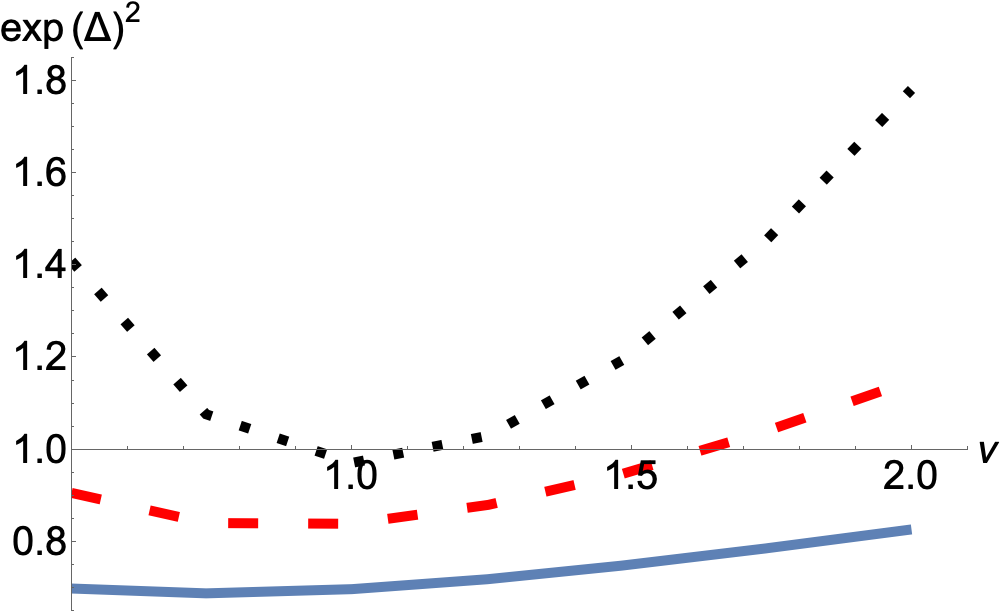}
  \caption{{$X_1 \sim \mathcal{N}\left(0,1\right)$, $\sqrt{6}\mathcal{E}_2 + 1 \sim \chi_1^2$}}
  \label{fig:ga-xi-pot}
 \end{subfigure}
\begin{subfigure}{.33\textwidth}
  \centering
  \includegraphics[width=.8\linewidth]{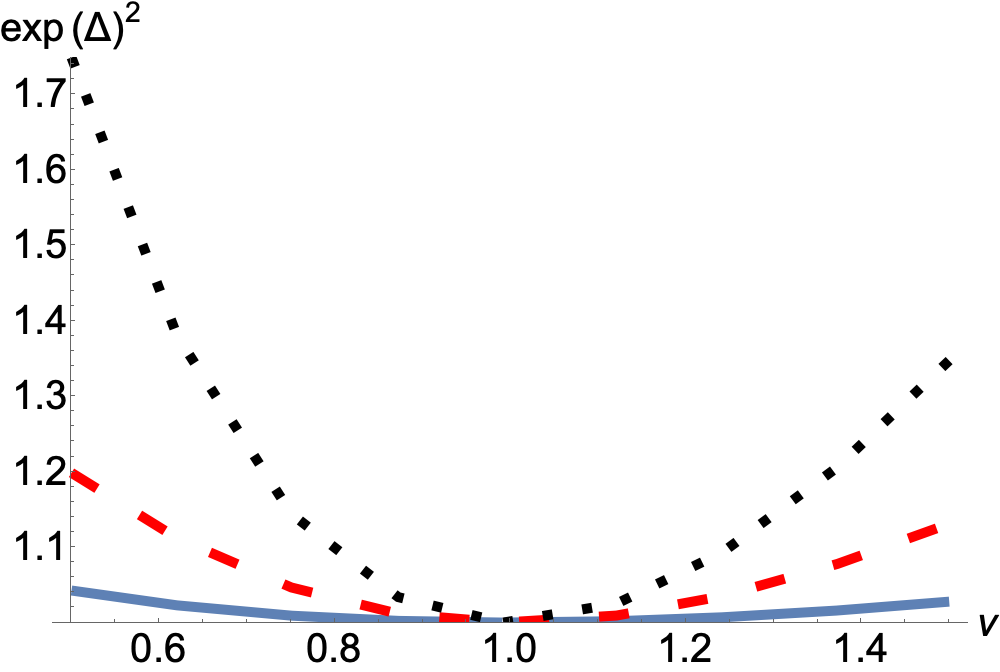}
  \caption{{$X_1 \overset{\mathcal{D}}{=} \sqrt{3}\mathcal{E}_2\sim \mathcal{N}\left(0,1\right)$}}
  \label{fig:ga-ga-pot}
\end{subfigure}
\caption{Two variable nonlinear model \eqref{eq:potmod}: effect of changing $\nu$ for $\beta = 0.5$ (solid blue curve), $\beta = 1$ (dashed red curve), and $\beta = 2$ (dotted black curve).} 
\label{fig:pot}
\end{figure}

Consider the case $X_1 \sim \mathcal{N}\left(0,1\right)$ and $\mathcal{E}_2 \sim \text{Unif}\left[-1,1\right]$. Analytic expressions for $\EE\left[X_1 \vert X_2 \right]$ and $\text{Var}\left(X_1 \vert X_2\right)$ can be found in terms of the Gaussian cumulative distribution function and the $\Gamma$-function. For $\left(\sigma_1^{G^\perp}\right)^2 = \EE\left[\text{Var}\left(X_1 \vert X_2\right)\right]$ we invoke numerical integration. The obtained $\exp\left(\Delta\right)^2$ is shown in Figure \ref{fig:ga-uni-pot} for different values of $\beta$. It confirms the intuition, that sufficiently strong nonlinearity leads to $\Delta > 0$ even for non-Gaussian errors. For $\beta = 1$ and $\beta = 2$, the model becomes identifiable for most $\nu \neq 1$. For $\beta = 0.5$, stronger nonlinearities are necessary. Also, the minimum $\Delta$ is not obtained for $\nu = 1$ but at $\nu \approx 1.32$. Thus, not every nonlinear model is better identifiable than the corresponding linear model.

As in the linear case, we consider the effect of an assymetric error distribution, namely, a scaled and centered chi-squared distribution. We show this in Figure \ref{fig:ga-xi-pot}. The factor $\sqrt{6}$ leads to the different lines corresponding to the same respective signal-to-noise ratios. As expected, higher degrees of nonlinearity are necessary to obtain a positive $\Delta$.

We show the behavior for a correctly specified model in \ref{fig:ga-ga-pot}. For $\nu = 1$, $\Delta = 0$ as implied by the unidentifiability result, otherwise, $\Delta > 0$. For $\nu \neq 1$, higher signal-to-noise ratios lead to more distinct $\Delta > 0$.
\paragraph*{Monotonicity of $f_2\left(\cdot\right)$}
The nonlinearities discussed here are designed to be slight deviations from the linear model and, thus, chosen to be strictly monotone. Notably, for non-monotone functions, the intuition that the anti-causal model is harder to fit is more applicable. In particular, if $X_1$ is centered and symmetric, and $f_2\left(\cdot\right)$ is an even function, it holds $\EE\left[X_1\vert X_2 \right] \equiv 0$. Then,
\begin{equation}
\left(\sigma_1^{G^\perp}\right)^2= \text{Var}\left(X_1\right) \quad \text{and} \quad \exp\left(\Delta\right)^2=\dfrac{\text{Var}\left(X_1\right)\text{Var}\left(X_2\right)}{\text{Var}\left(X_1\right)\text{Var}\left(\mathcal{E}_2\right)} > 1.
\end{equation}
Thus, the gap condition \ref{ass:gap} is satisfied regardless of the distribution of $\mathcal{E}_2$ as long as $X_1$ and $X_2$ have finite variance.

\section{Heteroskedastic noise model}
A simple extension of model \eqref{eq:ANM}, that has recently gained some attention,  is the heteroskedastic noise model also referred to as location-scale noise model
\begin{equation*}
X_j \leftarrow f_j \left(X_{\text{PA}\left(j\right)}\right) + g_j \left(X_{\text{PA}\left(j\right)}\right)\mathcal{E}_j \quad \text{with} \quad \EE\left[\mathcal{E}_j\right]=0, \quad \EE\left[\mathcal{E}_j^2\right]= 1 \quad \forall j \in 1, \ldots, p,    
\end{equation*}
with some nonnegative functions $g_j \left(\cdot\right)$ \citep{strobl2022identifying, xu2022inferring, immer2022loci}. It comes with similar identifiability guarantees as the ANM when testing for mutual independence between the $\mathcal{E}_j$. Let accordingly
\begin{align*}
f^G_j\left(X_{\text{PA}^G\left(j\right)}\right) & =\EE\left[X_j \vert X_{\text{PA}^G\left(j\right)}\right], \quad  g^G_j\left(X_{\text{PA}^G\left(j\right)}\right)^2 = \EE\left[\left(X_j - f^G_j\left(X_{\text{PA}^G\left(j\right)}\right)\right)^2 \vert X_{\text{PA}^G\left(j\right)}\right] \quad \text{and} \\
\mathcal{E}_j^G & = \dfrac{X_j - f^G_j\left(X_{\text{PA}^G\left(j\right)}\right)}{g^G_j\left(X_{\text{PA}^G\left(j\right)}\right)}
\end{align*}
be the conditional means, conditional variances, and residuals according to any, potentially wrong, DAG $G$. Then, one gets
\begin{align*}
\mathcal{L}\left(G\right) & = \sum_{j=1}^p \EE\left[\log\left(\dfrac{p_j^G\left(\mathcal{E}_j^G\right)}{g^G_j\left(X_{\text{PA}^G\left(j\right)}\right)}\right)\right]\\
\mathcal{L}^\mathcal{N}\left(G\right) &\coloneqq -\sum_{j=1}^p \EE\left[\log\left(g^G_j\left(X_{\text{PA}^G\left(j\right)}\right)\right)\right] + C = -\sum_{j=1}^p \dfrac{1}{2} \EE\left[\log\left(g^G_j\left(X_{\text{PA}^G\left(j\right)}\right)^2\right)\right] + C \\
& \geq  -\sum_{j=1}^p \dfrac{1}{2}\log\left(\EE\left[g^G_j\left(X_{\text{PA}^G\left(j\right)}\right)^2\right]\right) + C = -\sum_{j=1}^p \log\left(\sigma^G_j\right) + C.
\end{align*}
\begin{figure}[t!]
\begin{subfigure}{.5\textwidth}
  \centering
  \includegraphics[width=.8\linewidth]{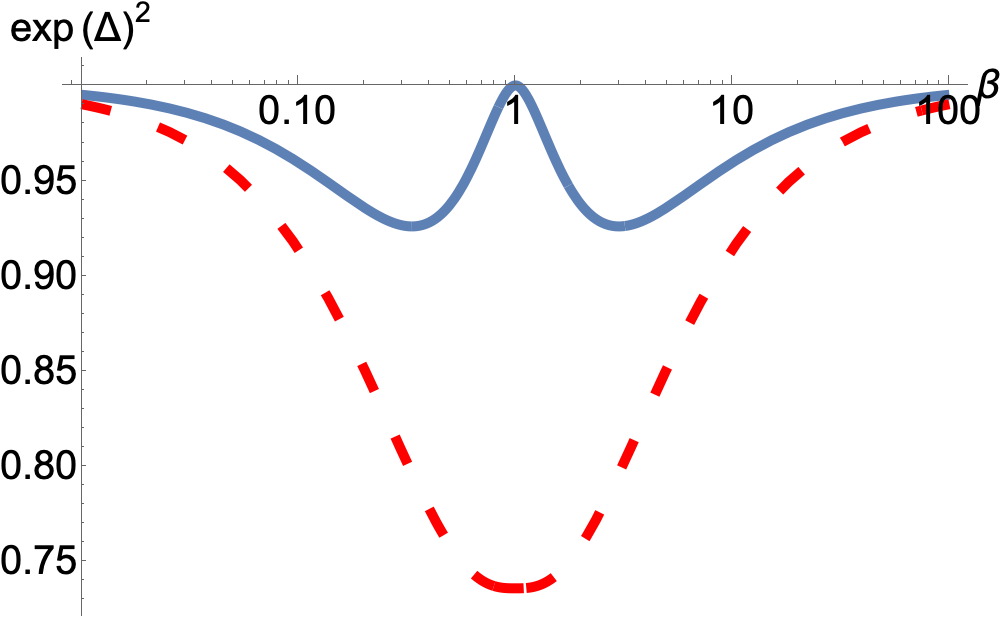}
  \caption{{Model \eqref{eq:linmod} with $X_1 \overset{\mathcal{D}}{=}\mathcal{E}_2 \sim \text{Unif}\left[-1,1\right]$}}
  \label{fig:uni-uni-het}
\end{subfigure}
\begin{subfigure}{.5\textwidth}
  \centering
  \includegraphics[width=.8\linewidth]{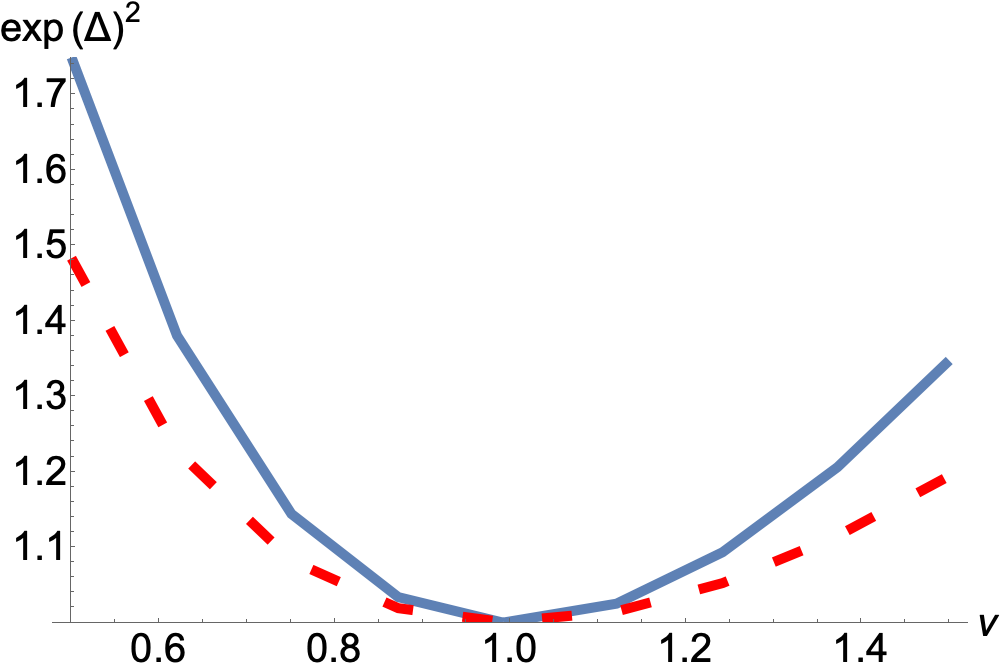}
  \caption{{Model \eqref{eq:potmod} with $X_1 \overset{\mathcal{D}}{=} \sqrt{3}\mathcal{E}_2\sim \mathcal{N}\left(0,1\right)$ and $\beta = 2$}}
  \label{fig:ga-ga-het}
\end{subfigure}
\caption{Two variable additive model: fitting homoskedastic models (solid blue curve) versus heteroskedastic models (dashed red curve).}
\label{fig:het}
\end{figure}
Thus, when fitting heteroskedastic models the score can only be increased compared to the homoskedastic fit. This can further increase the difficulty of finding the correct direction under non-Gaussian noise. Even if the true forward model is homoskedastic, i.e., $g_j\left(\cdot\right)  \equiv \sigma_j$, the backward model is typically heteroskedastic and profits from this new score. For example, in the set-up of Figure \ref{fig:uni-uni}, $\Delta$ would be negative even for $\beta = 1$. If one allows to fit heteroskedastic models, a result analagous to Theorem \ref{theo:lingam} exists. A negative gap is obtained unless all conditional expectations are linear and all conditional variances are constant for the wrong causal order.

In Figure \ref{fig:het}, we review the examples from Figures \ref{fig:uni-uni} and \ref{fig:ga-ga-pot} and see how allowing for a heteroskedastic fit makes the problem harder. For the sake of comparison, we look at $\exp\left(\Delta\right)^2$ although it does not have the same simple interpretation in the heteroskedastic case.

In terms of the location-scale noise model, the data-generating model as in Figure \ref{fig:uni-uni-het} is unidentifiable 
as $X_1\vert X_2$ is uniformly distributed, i.e., its distribution is independent of $X_2$ apart from location and scale. This does not contradict the identifiability results as they are derived for random variables with full support in $\mathbb{R}$.
\section{Discussion}
\subsection{Data applications}\label{dat}
For an extensive comparison between methods relying on Gaussian scoring and nonparametric independence tests in additive noise models or heteroskedastic noise models, we refer to \cite{immer2022loci}. There, several fitting methods are considered and combined with both approaches and evaluated on a variety of benchmark cause and effect pairs. Those pairs include both real and artificial data. For some of the considered data sources, using independence tests clearly improved the success rate for inferring the causal direction as compared to using the Gaussian score.

\begin{figure}[t!]
\centering
  \includegraphics[width=.9\linewidth]{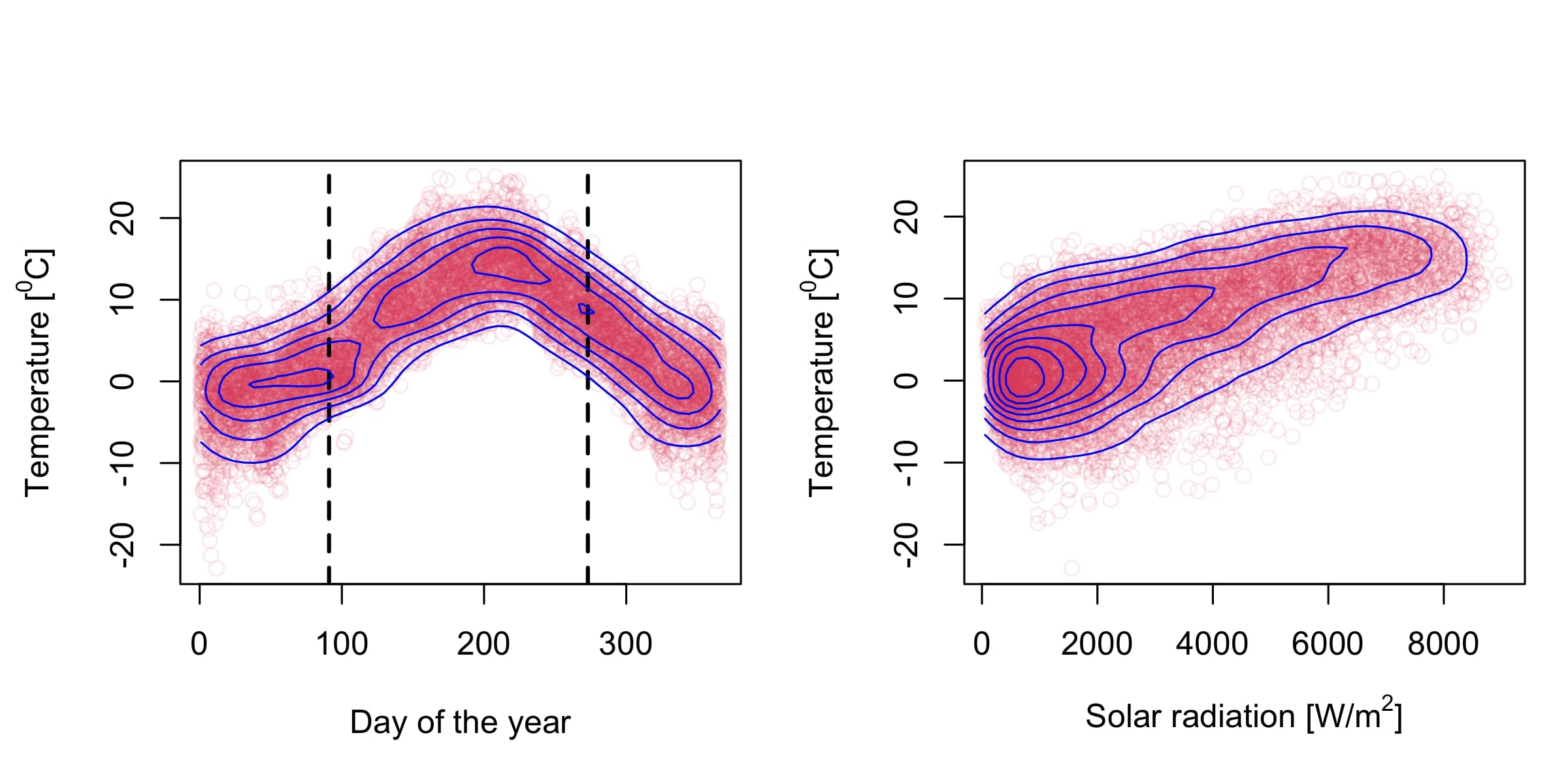}
  \caption{Scatter plot and contour lines of the density estimate for two selected pairs from the T\"ubingen data.}
  \label{fig:pairs}
\end{figure}

Let us consider two specific examples of the T\"ubingen data by \cite{mooij2016distinguishing}. Details on the data can be found in Section D.11 of their paper. Both examples have the temperature as the effect variable while the cause is the day of the year or the intensity of the solar radiation, respectively. We show the corresponding scatter plots as well as the contour lines of the density estimates in Figure \ref{fig:pairs}. It is evident that in neither case the cause variable is normally distributed, the days are perfectly uniformly distributed while solar radiation is right-skewed. Therefore, the assumptions for Gaussian scoring to infer the true causal direction are not fulfilled. For the first data set, we restrict the numerical analysis to the time frame April 1st to September 30th (March 31st to September 29th in leap years) to circumvent the issue that the data are circular. This is indicated by the black dotted lines.

To evaluate the Gaussian scores, we estimate the conditional expectation for either direction with a smoothing spline. In the first case, the causal effect is non-monotone, and the conditional mean in the anti-causal direction is not very informative to predict the day of the year. Therefore, we obtain the correct causal direction with Gaussian scoring even though the assumptions are not fulfilled. We get the data estimate $\exp\left(\Delta\right)^2 = 1.48$.

The effect of solar radiation on the temperature appears to be monotone which makes the conditional expectation in the anti-causal direction more informative. Also, it seems that the conditional expectation in the causal direction is not so far from being linear. This indeed makes the Gaussian scoring algorithm prefer the wrong direction. The estimate is $\exp\left(\Delta\right)^2 = 0.91$. Similarly, it fails for the first data set when considering the first 183 days of the year instead ($\exp\left(\Delta\right)^2 = 0.99$).

With RESIT relying on independence testing, we see for both data sets that the hypothesis of residuals being independent of the predictor is rejected in either direction. This indicates that the ANM in \eqref{eq:ANM} is not rich enough to explain the data. However, applying Algorithm 1 from \cite{peters2014causal} which minimizes the estimated dependence between predictor and residuals finds the true causal direction for both data pairs.

\subsection{Conclusion}
We discuss causal discovery in structural causal models using Gaussian likelihood scoring and analyze the effect of model misspecification. 

In the case where the data-generating distribution comes from a linear structural equation model and linear regression functions are used for estimation, the following holds. When the true error distribution is Gaussian, one can only identify the Markov equivalence class of the underlying data-generating DAG. The same is true when the error distribution is non-Gaussian but one wrongly relies on a Gaussian error distribution for estimation.

Thus, popular algorithms like the greedy equivalence search (GES) \citep{chickering2002optimal} for Gaussian models or the PC algorithm \citep{spirtes2000causation} assessing partial correlation are potentially conservative and only infer the Markov equivalence class when the error distributions are non-Gaussian as they do not exploit the maximal amount of information. But they are safe to use within the domain of data-generating linear structural equation models.
We prove here that this fact does not necessarily hold true when invoking nonparametric regression estimation.
Especially, if the true causal model is linear or just ``slightly nonlinear'' one would systematically get the wrong causal direction under error misspecification. As optimizing Gaussian scores is the same as optimizing $\ell_2$-loss, regressors that are more flexible than necessary for the causal model support anti-causal decisions. The intuition carries over when allowing for the flexibility of heteroskedastic error terms. If the true causal model has homoskedastic additive errors, fitting heteroskedastic models will increase the range of set-ups where misspecified Gaussian scoring chooses anti-causal relationships.

To overcome these issues, one could rely on general nonparametric independence tests, either between the different residuals or between residuals and predictors. Of course, this comes at higher computational cost and potentially lower sample efficiency in cases where Gaussian scoring works, including in the presence of non-monotonic causal effects.\\

\noindent \textbf{Acknowledgment:} The project leading to this application has received funding from the European Research Council (ERC) under the European Union’s Horizon 2020 research and innovation programme (grant agreement No 786461).\\

\noindent \textbf{Conflict of interest:} Authors state no conflict of interest.\\

\noindent \textbf{Data availability statement:}
The datasets analysed during the current study are available in the database with cause-effect pairs, \href{https://webdav.tuebingen.mpg.de/cause-effect/}{https://webdav.tuebingen.mpg.de/cause-effect/}. We consider pairs 42 and 77.

\bibliographystyle{apalike} 
\bibliography{references_gs}

\newpage
\appendix
\allowdisplaybreaks
\section{Proofs}
\subsection{Proof of Theorem \ref{theo:lingam}} 
 It is well known that for jointly Gaussian variables $X_1,\ldots ,X_p$ every possible causal order can induce the multivariate distribution with a suitable linear model. As the sum of log-variances determines the Kullback-Leibler divergence in the Gaussian case, this sum must be equal for all these linear models that induce the same multivariate distribution.

For every possible multivariate distribution with existing and bounded moment matrix $\boldsymbol{\Sigma}^\mathbf{X}=\EE\left[\mathbf{X}\mathbf{X}^\top\right]$, which the assumed model has, the linear least squares parameter and corresponding residual variances using arbitrary sets of regressor covariates are completely determined by the moment matrix. Therefore, the residual variances correspond to those of multivariate Gaussian data. Accordingly, one can obtain the same sum of log-variances as for the true model using the best linear predictors for every possible permutation. This proves the non-strict inequality in the theorem.

If for some variable $j$ the conditional expectation given its parents (in the DAG $G^\pi$) is not a linear function, the linear model cannot be optimal in terms of residual variance. Therefore, $\sigma_j^{G^\pi}$ in an unrestricted model is lower than the residual standard error of the best fitting linear model, and the score is further improved. Hence, the inequality is strict as soon as there exists at least one such variable.

\subsection{Proof of Proposition \ref{prop:lin-unif}}\label{app:prop}
The variances of $X_1$ and $X_2$ as well as $\mathcal{E}_ 2$ follow directly from the properties of the uniform distribution.
\begin{align*}
\left(\sigma_1^{G^0}\right)^2 & =\text{Var}\left(X_1\right) = \dfrac{1}{3}, \quad \left(\sigma_2^{G^0}\right)^2=\text{Var}\left(\mathcal{E}_2\right) = \dfrac{1}{3}, \quad \text{and} \\
\left(\sigma_2^{G^\perp}\right)^2 & =\text{Var}\left(X_2\right) = \beta^2 \text{Var}\left(X_1\right) + \text{Var}\left(\mathcal{E}_2\right) = \dfrac{1}{3} \left(\beta^2 + 1\right).
\end{align*}
The last term requires some more work. Due to the symmetry, we can assume without loss of generality that $\beta \geq 0$. For the densities, we get
\begin{align*}
f_{X_1}\left(x_1\right) & =\dfrac{1}{2}\mathbbm{1}_{\left\{\left\vert x_1\right \vert \leq 1\right\}}, \quad f_{X_2\vert X_1}\left(x_2 ,x_1\right) = \dfrac{1}{2} \mathbbm{1}_{\left\{\left\vert x_2- \beta x_1\right \vert \leq 1\right\}} \quad \text{and}\\
f_{X_2}\left(x_2\right) & = \int \dfrac{1}{4} \mathbbm{1}_{\left\{\left\vert x_1\right \vert \leq 1\right\}} \mathbbm{1}_{\left\{\left\vert x_2- \beta x_1\right \vert \leq 1\right\}} dx_1 = \dfrac{1}{4}\left(\text{min}\left\{1, \dfrac{x_2+1}{\beta}\right\} - \text{max}\left\{-1, \dfrac{x_2-1}{\beta}\right\}\right)\coloneqq\dfrac{1}{4}\left(a - b\right).
\end{align*}
For notational simplicity, we define random variables $A$ and $B$ with realizations $a$ and $b$. We obtain the moments
\begin{align*}
\EE\left[X_1 \vert X_2\right] & = \int x_1 f_{X_1\vert X_2} \left(x_1 , X_2 \right)dx_1 = \dfrac{\int  \dfrac{x_1}{4} \mathbbm{1}_{\left\{\left\vert x_1\right \vert \leq 1\right\}} \mathbbm{1}_{\left\{\left\vert X_2- \beta x_1\right \vert \leq 1\right\}} dx_1}{\int \dfrac{1}{4} \mathbbm{1}_{\left\{\left\vert x_1\right \vert \leq 1\right\}} \mathbbm{1}_{\left\{\left\vert X_2- \beta x_1\right \vert \leq 1\right\}} dx_1} = \dfrac{A^2 - B^2}{2 \left(A - B\right)} = \dfrac{1}{2}\left(A + B\right), \\
\EE\left[X_1^2 \vert X_2\right] & = \int x_1^2 f_{X_1\vert X_2} \left(x_1 , X_2 \right)dx_1 = \dfrac{\int  \dfrac{x_1^2}{4} \mathbbm{1}_{\left\{\left\vert x_1\right \vert \leq 1\right\}} \mathbbm{1}_{\left\{\left\vert X_2- \beta x_1\right \vert \leq 1\right\}} dx_1}{\int \dfrac{1}{4} \mathbbm{1}_{\left\{\left\vert x_1\right \vert \leq 1\right\}} \mathbbm{1}_{\left\{\left\vert X_2- \beta x_1\right \vert \leq 1\right\}} dx_1} = \dfrac{A^3 - B^3}{3 \left(A-B\right)} \\
& = \dfrac{1}{3}\left(A^2 + AB + B^2\right), \\
\text{Var}\left(X_1 \vert X_2\right) & = \EE\left[X_1^2 \vert X_2\right] - \EE\left[X_1 \vert X_2\right]^2 = \dfrac{1}{12} \left(A-B\right)^2.
\end{align*}
Finally, we are interested in
\begin{align*}
\left(\sigma_1^{G^\perp}\right)^2 & = \EE\left[\text{Var}\left(X_1 \vert X_2\right)\right] = \EE\left[\dfrac{1}{12} \left(A-B\right)^2\right] = \int_{-1 -\beta}^{1 + \beta} \dfrac{1}{12} \left(a-b\right)^2 f_{X_2}\left(x_2\right) dx_2\\
& = \int_{-1 -\beta}^{1 + \beta} \dfrac{1}{48} \left(a-b\right)^3 dx_2 = 2 \int_{0}^{1 + \beta} \dfrac{1}{48} \left(a-b\right)^3 dx_2.
\end{align*}
Assume first $\beta \geq 1$. Then, $b =\left(x_2 -1 \right)/\beta$ in the area of integration. For $x_2 \geq \beta - 1$, it holds $a = 1$.
\begin{align*}
\left(\sigma_1^{G^\perp}\right)^2 & = \int_{0}^{1 + \beta} \dfrac{1}{24} \left(a-b\right)^3 dx_2 = \int_{0}^{\beta -1} \dfrac{1}{24} \left(\dfrac{2}{\beta}\right)^3 dx_2 + \int_{\beta - 1}^{1 + \beta} \dfrac{1}{24} \left( \dfrac{\beta + 1 -x_2}{\beta}\right)^3 dx_2 \\
& = \dfrac{1}{24 \beta^3}\left(8 \left(\beta -1\right) + \int_0^2 u^3 du \right) = \dfrac{1}{24 \beta^3}\left(8 \beta - 4 \right) = \dfrac{1}{6 \beta^3}\left(2 \beta - 1 \right),
\end{align*}
where we applied the change of variable $u = \beta + 1 - x_2$ to simplify the integration.
Inserting residual variances with $\gamma = \beta$, the proposition's statement follows.

Alternatively, if $\beta < 1$, $a=1$ in the interval of integration. For $x_2 < 1 - \beta$, it holds $b=-1$.
\begin{align*}
\left(\sigma_1^{G^\perp}\right)^2 & = \int_{0}^{1 + \beta} \dfrac{1}{24} \left(a-b\right)^3 dx_2 = \int_{0}^{1 - \beta} \dfrac{1}{24} \left(2\right)^3 dx_2 + \int_{1- \beta}^{1 + \beta} \dfrac{1}{24} \left( \dfrac{\beta + 1 -x_2}{\beta}\right)^3 dx_2 \\
& = \dfrac{1}{24}\left(8 \left(1 - \beta\right) + \dfrac{1}{\beta^3}\int_0^{2\beta} u^3 du \right) = \dfrac{1}{24}\left(8 - 4\beta \right) = \dfrac{1}{6}\left(2  - \beta \right),
\end{align*}
where we applied the change of variable $u = \beta + 1 - x_2$ to simplify the integration.
Inserting all the residual variances with $\gamma = 1/\beta$, the proposition's statement follows.

\section{Derivations for the figures}
Assume model \eqref{eq:potmod}, which has model \eqref{eq:linmod} as a special case for $\nu = 1$. With $X_1 \sim \mathcal{N}\left(0,1\right)$ the normalization is
\begin{equation*}
V\left(\nu\right)\coloneqq\EE\left[\left\vert X_1\right\vert^{2\nu}\right] = \dfrac{2^\nu \Gamma\left(\nu + \dfrac{1}{2}\right)}{\sqrt{\pi}}.    
\end{equation*}
As before,
\begin{align*}
\left(\sigma_1^{G^0}\right)^2 & =\text{Var}\left(X_1\right), \quad \left(\sigma_2^{G^0}\right)^2=\text{Var}\left(\mathcal{E}_2\right) \quad \text{and} \\
\left(\sigma_2^{G^\perp}\right)^2 & =\text{Var}\left(X_2\right) = \beta^2 + \text{Var}\left(\mathcal{E}_2\right).
\end{align*}
\subsection{Gaussian and uniform}
For $\mathcal{E}_2 \sim \text{Unif}\left[-1,1\right]$, $X_1$ is constrained to lie within 
\begin{equation*}
\text{Sign}\left(\left(X_2-1\right)\sqrt{V\left(\nu\right)}/\beta\right)\left\vert \left(X_2-1\right)\sqrt{V\left(\nu\right)}/\beta \right\vert ^{1/\nu} \quad \text{and} \quad \text{Sign}\left(\left(X_2+1\right)\sqrt{V\left(\nu\right)}/\beta\right)\left\vert \left(X_2+1\right)\sqrt{V\left(\nu\right)}/\beta \right\vert ^{1/\nu},
\end{equation*}
i.e., given $X_2$, it is a truncated standard Gaussian. The conditional mean and variance follow from standard theory about truncated Gaussian random variables. For simplicity, call the upper bound $A$ and the lower bound $B$. The density of $X_2$ is then given by
\begin{equation*}
f_2\left(x_2\right) = \int_{b}^a \dfrac{1}{2} \phi\left(x_1\right) dx_1 = \dfrac{1}{2}\left(\Phi\left(a\right) - \Phi\left(b\right)\right).
\end{equation*}
Finally,
\begin{equation*}
\left(\sigma_1^{G^\perp}\right)^2 = \EE\left[\text{Var}\left(X_1 \vert X_2\right)\right]
\end{equation*}
is obtained by numerically integrating over (a sufficient part of) the real line.
\subsection{Two Gaussian random variables, or Gaussian and $\chi_1^2$}
Except for $V\left(\nu\right)$, all quantities are obtained by brute force numerical integration.
\subsection{Two uniform random variables with heteroskedastic fitting}
We can mainly follow the derivation in \ref{app:prop}. Instead of $\left(\sigma_1^{G^\perp}\right)^2$, we need 
$\exp\left(\EE\left[\log\left(\text{Var}\left(X_1 \vert X_2\right)\right)\right]\right)$
which is obtained by numerical integration.
\end{document}